\documentclass[12pt]{article}
\usepackage{amsmath}
\usepackage{amssymb}
 \usepackage{graphicx}
\usepackage{color}
\usepackage{url}
\newcommand*{\QED}{\hfill\ensuremath{\square}\vspace{0.5em}}%

\newtheorem{lemma}{Lemma}
\newtheorem{corollary}{Corollary}
\newtheorem{theorem}{Theorem}

\title{Counting and Enumerating Galled Networks}
\author{Andreas DM Gunawan\thanks{Department of Mathematics, National University of Singapore, 
Singapore 119076}, Jeyaram Rathin\thanks{Department of Applied Mathematics and Computational Sciences, PSG College of Technology, Coimbatore - 641004, India. This work was done when he visited  the National University of Singapore as an exchange student.}, Louxin Zhang\thanks{Department of Mathematics, National University of Singapore, 
Singapore 119076. Email: matzlx@nus.edu.sg}}

\begin{document}

\maketitle

\begin{abstract}
Galled trees are widely studied as a recombination model in population genetics.  This class of phylogenetic networks is generalized into galled networks by relaxing a structural condition.   In this work, a  linear recurrence  formula is given for counting 1-galled networks, which are galled networks satisfying the condition that each reticulate node has only one leaf descendant.  Since every galled network consists of a set of 1-galled networks stacked one on top of the other, a method is also presented to count and enumerate galled networks. 
\end{abstract}

\section{Introduction}
Phylogenetic networks have been used  more and more frequently  in evolutionary genomics and population genetics in the past two decades \cite{Gusfield_book, Lake_99}. A rooted phylogenetic network (RPN) is a rooted acyclic digraph in which  all the sink nodes are of indegree 1 and there is a unique source node called the root, where the former represent a set of taxa (e.g, species, genes, or individuals in a population) and the latter represents the least common ancestor of the taxa.  Moreover, RPNs also satisfy the property that non-leaf and non-root nodes are of either indegree 1 or outdegree 1; these nodes are called tree nodes and reticulate nodes, respectively. 

 Imposing  topological conditions on the network allows us to define different classes of RPNs such as galled trees \cite{Gusfield_04, Wang_01}, galled networks \cite{Huson_07}, tree-child networks \cite{Cardona_09b}, reticulation-visible networks \cite{Huson_book} and tree-based networks \cite{Francis_15,Zhang_16}
(see also \cite{Steel_book, Zhang_18}). A galled tree is a binary RPN such that (i)  for each reticulate node $u$, with the parents being denoted as $p'(u)$ and $p''(u)$, there are two 
edge-disjoint paths from the least common ancestor of $p'(u)$ and $p''(u)$ to $u$ that contain only tree nodes except for $u$,  and (ii) for any two reticulate nodes, the paths  in (i) do not overlap \cite{Wang_01}. Later, galled networks were defined by Huson and Kl{\"o}pper  to be RPNs that satisfy only Property (i)   
\cite{Huson_07}. Reconstruction of galled networks has also been studied in  \cite{Huson_09}.   
These network classes are of particular interest because  they have nice combinatorial properties. Moreover,   some important NP-complete problems related to phylogenetic trees and clusters can be solved in polynomial-time when restricted to these classes \cite{Bordewich_16, Gambette_15, Gunawan_16, Gunawan_17}.

In this paper,  we investigate how many galled networks exist over a set of taxa. Phylogenetic trees are RPNs without any reticulate node.
It is well known that $(2n-3)!!$ binary phylogenetic trees exist over $n$ taxa. However, counting becomes much harder for 
general RPNs. For example, even counting RPNs with a couple of reticulate nodes is challenging \cite{Fuchs_18}.
Recent advances in counting have been made for tree-child networks \cite{Fuchs_18,Semple_15} and galled trees \cite{Bouvel_18,Steel_06}.  Here, we provide a linear recurrence formula for finding  the number of 1-galled networks, which are galled networks such that each reticulate node has only one leaf descendant. The formula is obtained through a connection between galled networks and leaf-multi-labeled (LML)  trees.  Since  a galled network is essentially a set of 1-galled networks stacked one on top of the other in a tree-like structure, we also present a general method for counting and enumerating galled networks.  Although counting LML trees was investigated by Czabarka et al. \cite{Moulton_13}, our results are not  derived from their study. 

The rest of this paper is divided into five sections. Section~\ref{sec2} introduces some basic notation that are necessary for our study. 
Section~\ref{sec3} establishes the fact that 1-galled networks have  a one-to-one correspondence with the so-called dup-trees. Section~\ref{sect5} presents a linear recurrence formula for counting 1-galled networks. Section~\ref{sec5} examines how to count and enumerate general galled networks.  Section~\ref{sec6} concludes the study with a few remarks.

\section{Basic notation}
\label{sec2}

\subsection{Phylogenetic networks}

 A binary RPN over a finite set of taxa $X$ is an  acyclic digraph such that:
\vspace{-0.5em}
\begin{itemize}
   \item there is a unique node of indegree 0 and outdegree 2, called its {\it root}; \vspace{-0.5em}
   \item there are exactly $|X|$ nodes of indegree 1 and outdegree 0, called the {\it leaves} of the RPN, each labeled with a unique taxon in $X$; and \vspace{-0.5em}
 \item  each non-leaf/root node is either a {\it reticulate node} that is of indegree 2 and outdegree 1,  or 
a {\it tree node} of indegree 1 and outdegree 2.
\end{itemize}
Three RPNs are illustrated in Figure~\ref{Fig_1}, where each edge is directed away from the root and edge orientation is not omitted. 
  For a RPN $N$, we  use ${\cal V}(N)$ and ${\cal A}(N)$ to denote its node set and directed edge set, respectively. 

Let $u$ and $v$ be two nodes of $N$.  The node $u$ is said to be a {\it parent} (resp. a {\it child}) of  $v$ if $(u, v)\in {\cal A}(N)$ (resp. $(v, u)\in {\cal A}(N)$).  Each reticulate node $r$ has a unique child and we denote this as $c(r)$. Each tree node $t$ has a unique parent and we denote this as $p(t)$.
In general, $u$ is an {\it ancestor} of $v$ (or equivalently, $v$ is {\it below} $u$) if there is a direct path from the root of $N$ to $v$ that contains $u$. 

A binary {\it phylogenetic tree} over a set of taxa $X$  is simply a binary RPN containing no reticulate nodes.

A RPN is said to be {\it galled} if every reticulate node $r$ has an ancestor $a_r$ such that there are edge-disjoint paths
from $a_r$ to $r$ that do not contain any reticulate nodes other than $r$. The RPN in Figure~\ref{Fig_1}a is not galled but 
the one in Figure~\ref{Fig_1}b is. By definition, every RPN with only one reticulate node or without reticulate nodes is galled.

\subsection{Dup-trees}

In this work, we will count galled networks through the connection between galled networks and the so-called LML trees. 

\begin{figure}[t!]
            \centering
            \includegraphics[scale = 0.8]{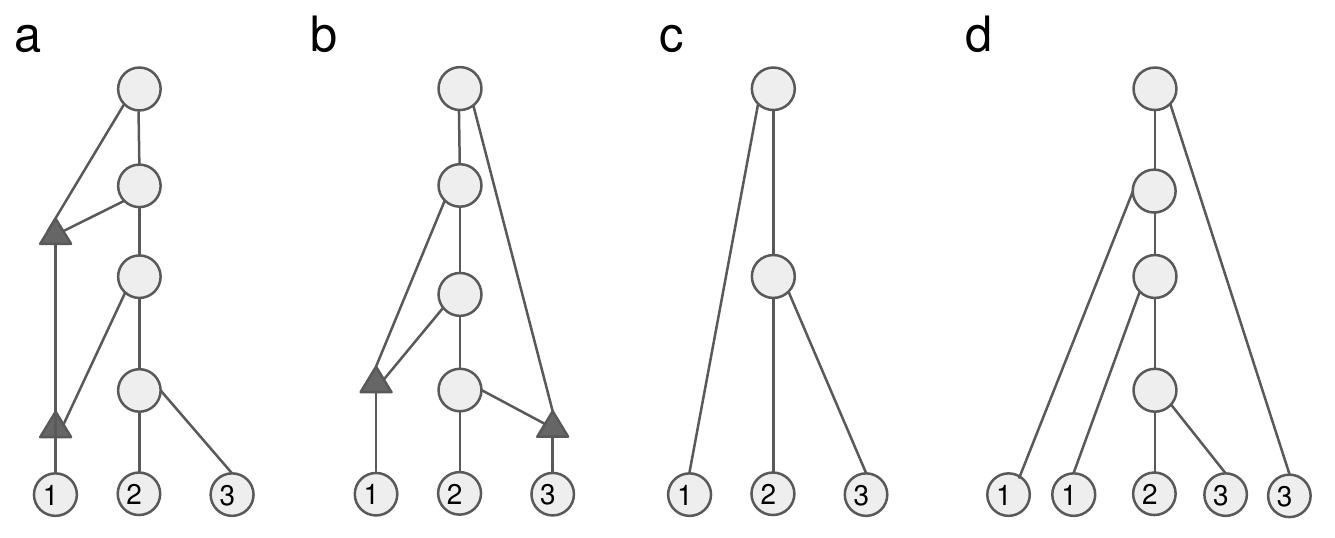}
           \caption{ RPNs and trees over $\{1, 2, 3\}$, where reticulate and tree nodes are drawn as filled triangles and open circles, respectively. ({\bf a}) A binary RPN.  ({\bf b}) A binary galled network. 
 ({\bf c})  A binary phylogenetic tree. 
 ({\bf d})  A rooted binary dup-tree, where the labels `1' and `3' are duplicated labels.}
            \label{Fig_1}
\end{figure}

A rooted (resp. unrooted) {\it  LML  tree} is a binary rooted (resp. unrooted)  tree with leaves that are labeled in a way such that several leaves may have an identical label.  
It is a {\it dup-tree} if at most two leaves have the identical label.  A rooted dup-tree is given in  Figure~\ref{Fig_1}d.  
Here,  a phylogenetic tree is considered to be  a trivial rooted dup-tree. The child--parent and ancestor--descendant relationships can be defined for nodes in a rooted dup-tree in the same way as in a RPN. 

Let $M$ be a dup-tree over $X$.  
A taxon $x\in X$ is said to be a {\it duplicated label} for $M$ if  two distinct leaves labeled with $x$ exist and a {\it 1-label} otherwise.  
$L_1(M)$ and $L_{2}(M)$  are used to denote the subsets of the  1-labels and duplicated labels in $M$, respectively.

A {\it cherry} in a dup-tree is a pair of leaves that are adjacent to a common non-leaf node. 
A cherry is said to be a {\it twin}-{\it cherry} if two leaves belonging to it are labeled with a common taxon. 
A dup-tree is said to be {\it twin-cherry-free}  if it does not contain any twin-cherries.

Let $M$ be a unrooted LML tree over $X$,  $x\in X$ and $e=(u, v)\in {\cal A}(M)$.  {\it Grafting} a new leaf $x$ to 
$e$ involves  replacing $e$ by a path consisting of two paths $(u, p)$ and $(p, v)$, and attaching the leaf
as the child of $p$, where $p$ is not in $M$.  Conversely,  for a leaf $\ell$ in $M$, its parent $p(\ell)$ is adjacent to 
two nodes $x$ and $y$ other than $\ell$. {\it Pruning} $\ell$ from $M$ means removing $\ell$ and $p(\ell)$  and any incident edges and then adding 
$(x, y)$ as an edge to $M$.   In this work, we use $M\oplus (e, x)$ to denote the tree obtained from grafting $x$ to $e$ in $M$, or $M\oplus x$ if there is no confusion if $e$ is omitted.  Similarly, $M\ominus \ell$ is used to denote the tree obtained from $M$ by pruning $\ell$ for a leaf $\ell$ in $M$. 

\subsection{Decomposition  of galled networks into tree-components}

Consider a RPN $N$. Let ${\cal R}(N)$ and ${\cal L}(N)$ denote the sets of reticulate nodes and leaves in $N$, respectively. 
The subnetwork 
$N-\left(\mathcal{R}(N)\cup \mathcal{L}(N)\right)$ is a forest for which each connected component consists of tree nodes. 
Each connected component  is called a {\it tree-component} of $N$ \cite{Gunawan_18, Zhang_18}.  Note that each tree-component does not contain any leaves. This is different from the definition of tree-components given in \cite{Gunawan_16}.

A reticulate node is {\it inner} if both its parents are in a common tree-component. It is a {\it cross} reticulate node otherwise.  Galled networks have the following recursive characterization. 

\begin{theorem} 
\label{thm1}
Let $G$ be a galled network.   \vspace{-0.5em}
\begin{itemize}
  \item[{\rm (1)}]  Each reticulate node is inner in $G$. \vspace{-0.5em}
  \item[{\rm (2)}]  For any $r\in \mathcal{R}(G)$, $G-\{r\}$ consists of two connected components, and  the component contains all the descendants of $r$ form a galled subnetwork rooted at the child $c(r)$ of $r$.
\end{itemize} 
\end{theorem}
{\bf Proof.}  (1) It has been proven \cite[Theorem 2]{Gunawan_16} that a binary RPN is galled if and only if every reticulate node is inner.

(2)  Clearly,  the statement  follows from Part 1. 
$\QED$

A RPN is a {\it 1-galled} network if it is a galled network with only one tree-component. The RPN in  
Figure~\ref{Fig_1}b is 1-galled.  It is easy to derive the following facts from  Theorem~\ref{thm1}.

\begin{corollary} Let $N$ be a RPN.   \vspace{-0.5em}
\begin{itemize}
  \item[{\rm (1)}]  If there is only one tree-component in $N$, then $N$ is galled. \vspace{-0.5em}
  \item[{\rm (2)}]  If every reticulate has only one leaf descendant in $N$, then $N$ is 1-galled.
\end{itemize} 
\end{corollary}

\section{Dup-trees and 1-galled networks}
\label{sec3}

Let $M$ be a dup-tree over $X$. Recall that $L_2(M)$ denotes the subset of duplicated labels. For each $x \in L_2(M)$, we use $\ell'(x)$ and $\ell''(x)$ to denotes the two leaves that are  labeled with $x$.

Let us assume that $M$ is twin-cherry-free. We derive a RPN ${\cal N}(M)$ by (i) removing $\ell'(x)$ and $\ell''(x)$,   (ii) introducing a reticulate node $r_x$,  (iii) connecting the parents $p(\ell'(x))$ and 
$p(\ell''(x))$ of $x$ to $r_x$,  and (iv) attaching a leaf $\ell_x$ with the label below $r_x$  for each duplicated label $x$.  Formally, 
${\cal N}(M)=(\bar{V}, \bar{A})$, where:
\begin{eqnarray}
   \bar{V}&=&\left[ \mathcal{V}(M) - \left\{\ell'(x), \ell''(x) \;|\; x\in L_2(M)\right\}\right] \cup 
\left\{r_x, \ell_x \;|\; x\in L_2(M)\right\},  \\
\bar{A}&=&\left[\mathcal{A}(M)- \left\{ (p(\ell'(x)), \ell'(x)),  (p(\ell''(x)), \ell''(x)) \;|\; x\in L_2(M)   \right\}\right] \nonumber  \\
 && \cup \left\{(p(\ell'(x)), r_x),  (p(\ell''(x)), r_x), (r_x, \ell_x)  \;|\; x\in  L_2(M)  \right\}.
\end{eqnarray}
If $M$ is a phylogenetic tree, $\mathcal{N}(M)$ is just $M$. If $M$ is a dup-tree containing at least one duplicated label, $\mathcal{N}(M)$ is then a 1-galled network containing as many reticulate nodes as the duplicated labels in $M$.  
This transformation from a LML tree to a network is called the ``folding" operation in \cite{Huber_06}.

Conversely, it is not hard to see that splitting each reticulate node in a 1-galled network $N$ results in a dup-tree $M$ such that 
$\mathcal{N}(M)=N$.  This proves the following statement: 

\begin{theorem}
\label{thm31}
 Let $X$ be a finite set and $r\geq 1$. There is a one-to-one correspondence between \vspace{-0.5em}
 \begin{itemize}
\item      The  binary  twin-cherry-free dup-trees with $r$ duplicated labels over $X$, and \vspace{-0.5em}
\item    The binary 1-galled networks with $r$ reticulate  nodes.
\end{itemize}
\end{theorem}

Note that the 1-galled network in Figure~\ref{Fig_1}b corresponds with the dup-tree in Figure~\ref{Fig_1}d.

\section{Counting 1-galled networks}
\label{sect5}

Without loss of generality, we set $[k]=\{1, 2, ..., k\}$. We adopt the following notation:\vspace{-0.5em}
 \begin{itemize}
  \item $\mathcal{T}_k$ is the set of phylogenetic trees over $[k]$. \vspace{-0.5em}
  \item $\mathcal{UT}_k$ is the set of  binary unrooted trees over $[k]$. \vspace{-0.5em}
   \item $\mathcal{D}_{i, k}$ is the set of rooted dup-trees $M$ over $[k]$ such that  $M$ is twin-cherry-free and 
    $L_2(M)=[i]$, where $1\leq i\leq k$. \vspace{-0.5em}
  \item $\mathcal{UD}_{i, k}$ is the set of  unrooted dup-trees $M$ over $[k]$ such that $M$ is twin-cherry-free and
   and  $L_2(M)=[i]$, where $1\leq i<k$. \vspace{-0.5em}
  \item $\mathcal{G}_{i, k}$ is the set of 1-galled networks over $[k]$ that has exactly $i$ reticulate nodes with the child being labeled with a unique element in $[i]$. 
\end{itemize}

\begin{lemma} 
\label{lemma1}
For any $k\geq 1$, 
 \begin{eqnarray} 
  &&|\mathcal{T}_k|=|\mathcal{UT}_{k+1}|=(2k-3)!!=(2k-3)\times (2k-5) \times \cdots \times  3\times 1, \\
  && |\mathcal{G}_{i, k}|= |\mathcal{D}_{i, k}|=|\mathcal{UD}_{i, k+1}|, \;\;i\leq k.
\end{eqnarray}
\end{lemma}
{\bf Proof.}  The first equation is well known (see  \cite[page 16]{Steel_book}). Similarly, by Theorem~\ref{thm31},  the second equation is also true. 
$\QED$

\subsection{A recursive formula for $|\mathcal{UD}_{i, k}|$ and $|\mathcal{G}_{i, k}|$}

 It is well known that 
every binary unrooted tree over $[k+1]$ can be obtained from a unique binary unrooted tree over $[k]$ by inserting the leaf labeled with $(k+1)$ on an edge of the latter.   In the section, we generalize this fact to give a recurrence formula for $|\mathcal{UD}_{i, k}|$ and $|\mathcal{G}_{i, k}|$. 

\begin{figure}[t!]
            \centering
            \includegraphics[scale = 1.0]{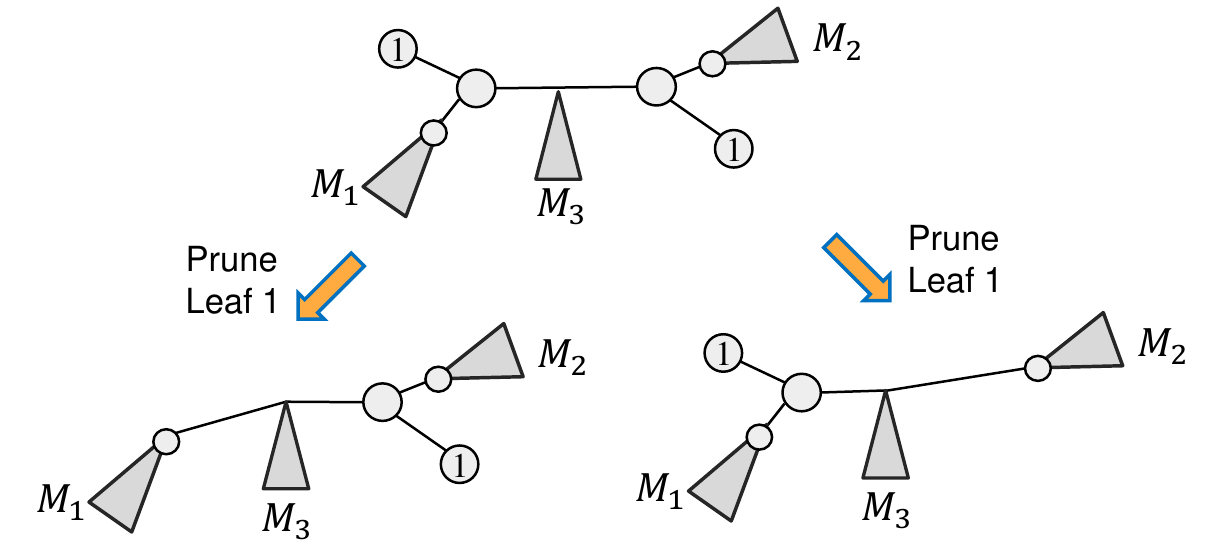}
           \caption{A dup-tree $M$ in $\mathcal{UD}_{1, k}$ can be partitioned into the three subtrees $M_1, M_2, M_3$, where
only $M_2$ can be empty.  Pruning different leaves labeled 1 from $M$ results in  two distinct trees in $\mathcal{UD}_{0, k}$ if $M_1$ is non-empty,  and an identical tree otherwise. }
            \label{fig2}
\end{figure}

As a warmup, we first count the dup-trees in $\mathcal{UD}_{1, k}$ for $k\geq 2$. For simplicity, we set 
$\mathcal{UD}_{0, k}=\mathcal{UT}_k$.  Let $M\in \mathcal{UD}_{1, k}$. Since $M$ is twin-cherry-free,  the leaves labeled with 1 are not sibling and thus $M$ can be partitioned into three parts ($M_1, M_2, M_3$) as illustrated in Figure~\ref{fig2}. 
 Pruning different leaves labeled with 1 from $M$ results in different trees in $\mathcal{UD}_{0, k}$ if the middle subtree $M_3$ is not empty,  and the same tree otherwise. 
This suggests that grafting an extra leaf labeled with 1 into every edge in each tree in $\mathcal{UD}_{0, k}$ can generate every tree
in $\mathcal{UD}_{1, k}$ twice. Note that if we graft the extra leaf into the edge incident to the original leaf labeled with 1, we get a dup-tree in which two leaves of label 1 form a twin-cherry, which is not in $\mathcal{UD}_{1, k}$.  

Since there are $2k-4$ edges that are not incident to the leaf labeled 1 in every unrooted binary tree in $\mathcal{UD}_{0, k}$, we have: 
 \begin{eqnarray*}
     2|\mathcal{UD}_{1, k}|=(2k-4) \cdot |\mathcal{UD}_{0, k}|.
\end{eqnarray*}
Therefore, by Lemma~\ref{lemma1}, we obtain:
 \begin{eqnarray}
     |\mathcal{UD}_{1, k}|=(k-2) \cdot (2k-5)!!.
\end{eqnarray}

In the rest of this section, we will focus on the case where  $i>1$.  The analysis for this case is more subtle than what we have done so far. 
Let $M\in  \mathcal{UD}_{i, k}$, where $i>1$. 

\begin{figure}[t!]
            \centering
            \includegraphics[scale = 0.8]{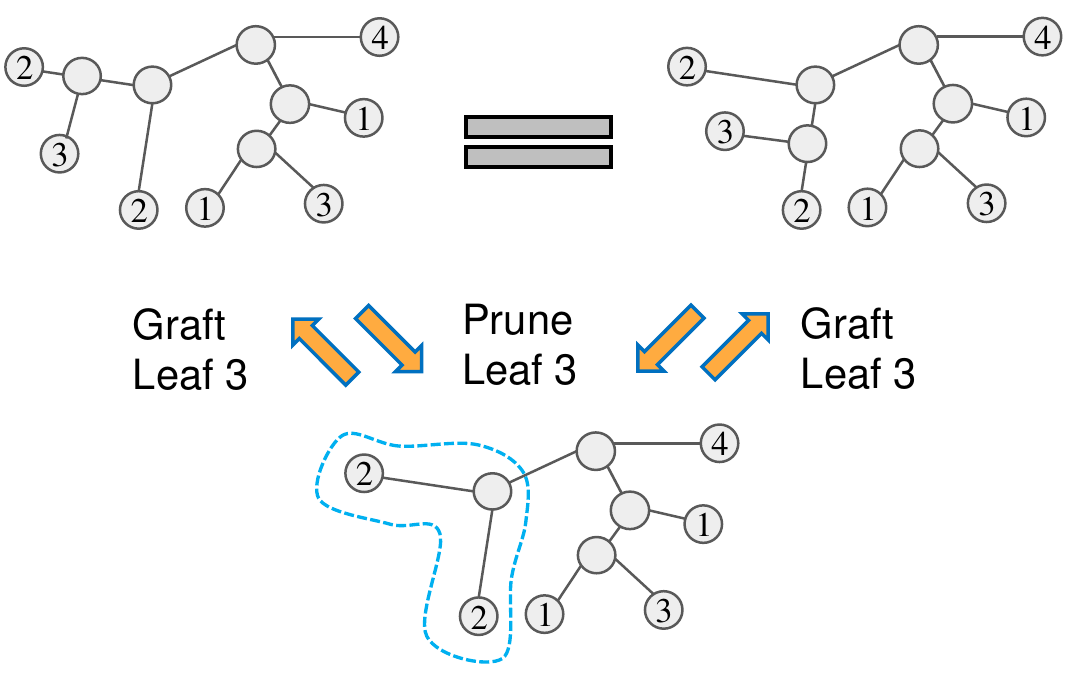}
           \caption{Grafting the second copy of Leaf 3 into either edge in the unique twin-cherry (circled) in a dup-tree $T$ (bottom) produces the same dup-tree (top), where $L_2(T)=\{1, 2\}$. }
            \label{fig_3}
\end{figure}

First, 
we have to graft a leaf labeled with $i$ 
into a twin-cherry in a dup-tree $T$ over $[k]$ such that $L_2(T)=[i-1]$ to get some dup-tree in $\mathcal{UD}_{i, k}$
 as illustrated Figure~\ref{fig_3}. In the dup-tree on the top in Figure~\ref{fig_3},  a leaf labeled with 3 is in the 
twin-cherry consisting of leaves labeled 1, whereas another is in the twin-cherry consisting of leaves labeled with 2.  In this case, we have the following fact.

\begin{lemma} 
\label{lemma2}
  Let $T$ be a dup-tree over $[k]$ such that $L_2(T)=[i-1]$, $i\leq k$. If $T$ contains  a unique  twin-cherry, then grafting a leaf labeled with  
$i$ into either edge  in the twin-cherry will produce the same tree in $\mathcal{UD}_{i, k}$.
\end{lemma}

Conversely, consider a unrooted dup-tree $M\in \mathcal{UD}_{i, k}$. For a non-leaf node $u$ and 
a node $v$ that is adjacent to $u$, we use $M_u(v)$ to denote the connected component containing $v$ in $M-u$ and call it a subtree adjacent to $u$.   The node $u\in {\cal V}(M)$ is said to be a {\it duplication node} if  it is adjacent to two nodes $v'$ and $v''$  such that 
 $M_u(v')$ and $M_u(v'')$ are identical as rooted trees;  in other words, there is a mapping 
$$f: \mathcal{V}(M_u(v')) \rightarrow  \mathcal{V}(M_u(v''))$$ such that (i) $f(v')=v''$,   (ii) $(x, y)\in  \mathcal{A}(M_u(v'))$ if and  only if  $(f(x), f(y)) \in  \mathcal{A}(M_u(v''))$ and (iii) $x$ is a leaf if and only if  
$f(x)$ is a leaf labeled with the same taxa, where  $x, y\in \mathcal{V}(M_u(v'))$.  
$M_u(v')$ and $M_u(v'')$ are called the {\it conjugate subtrees} of $u$ if $u$ is a duplication node. The two edges that are correspondent  with each other under $f$ are also said to be {\it conjugate}.

\begin{lemma} 
Let $M$ be a dup-tree that may contain  twin-cherries  and $L_2(M)=[i]$.   \vspace{-0.5em}
\begin{itemize}
  \item[{\rm (1)}]  The non-leaf node in a twin-cherry is a duplication node. \vspace{-0.5em}
   \item[{\rm (2)}] For different duplication nodes $u$ and $v$ in $M$, their conjugate subtrees are disjoint. 
\end{itemize}
\end{lemma}
{\bf Proof.}  (1) This derives from the definition of a duplication node. (2) For different duplication nodes $u$ and $v$, 
the conjugate subtrees associated with $u$ contain leaves with labels that are different from the labels appearing in the conjugate subtrees associated with $v$, as each duplicated element labels exactly two leaves.  
$\QED$

 Lemma~\ref{lemma2} can now  be generalized as follows.
\begin{lemma} 
\label{lemma4}
Let $M\in \mathcal{UD}_{i, k}$, where $i\leq k$,  and let  $u$ be a duplication node of $M$ with the conjugate subtrees  $M'$ and $M''$.  Grafting the second leaf labeled with $i+1$ into an edge $e$ in $M'$ will produce the same tree as grafting the leaf in the edge conjugate to $e$ in $M''$. 
\end{lemma}

Second, some unrooted dup-trees in $\mathcal{UD}_{i+1, k}$ are generated by grafting a new leaf labeled with $i+1$ in a dup-tree in three or four times. 
Specifically, we have the following fact: 

\begin{lemma}
\label{lemma5}
  Let $M\in \mathcal{UD}_{i+1, k}$, $i< k$. 
  \begin{itemize}
   \item[\rm{(1)}] If $\ell'(i+1)$ and $\ell''(i+1)$ are in the conjugate subtrees of 
a duplicate node in $M$,  then, $M\ominus \ell'(i+1)=M\ominus \ell''(i+1)$, from which  
 $M$ can only be generated by grafting a leaf labeled with $i+1$ in a unique edge.
   \item[{\rm (2)}] If neither $\ell'(i+1)$ nor  $\ell''(i+1)$ is in the conjugate subtrees of any duplicate  node,  
  $M\ominus \ell'(i+1)$ and $M\ominus \ell''(i+1)$ are different dup-trees in 
  $\mathcal{UD}_{i, k}$ if and only if  $p(\ell'(x))$ and $p(\ell''(x))$ are not adjacent. 
   \item[\rm{(3)}] If  $\ell'(i+1)$ is not in a conjugate  subtree of any duplication node and 
   if pruning $\ell'(i+1)$ from $M$ does not produce a new duplication node, then
     $M$ can be obtained from $M\ominus \ell'(i+1)$ by grafting a leaf labeled with $i+1$ in a unique edge.
    \item[\rm{(4)}] If   $\ell'(i+1)$ is not in a duplication subtree but
   $M\ominus \ell'(i+1)$  contains one duplication node that is not a duplication node in $M$,  
  then,  $M$ can be obtained by grafting a leaf labeled $i+1$ in two different edges 
  in $M\ominus \ell'(i+1)$.
\end{itemize} 
\end{lemma}

\begin{figure}[t!]
            \centering
            \includegraphics[scale = 0.8]{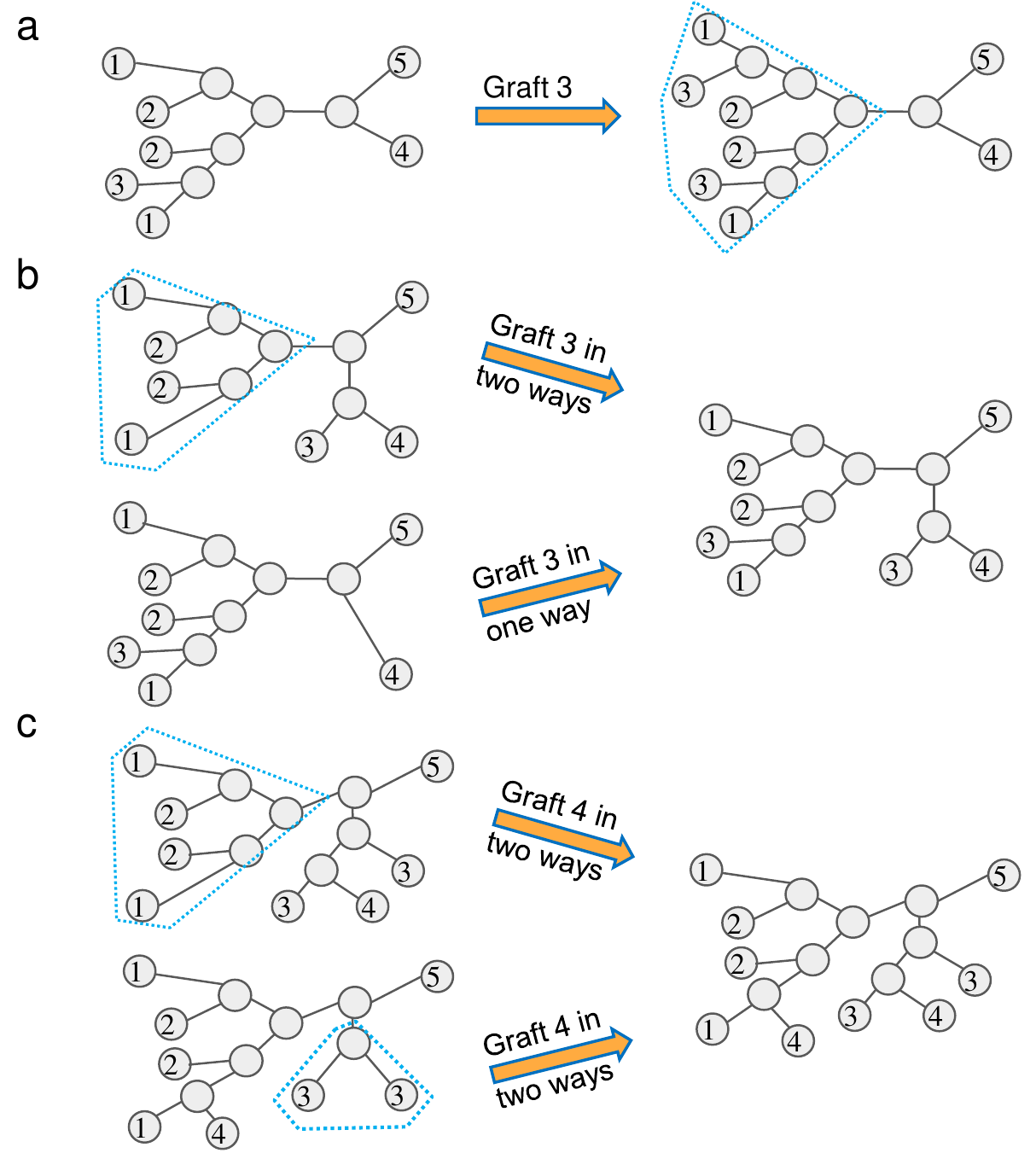}
           \caption{A unrooted dup-tree in $\mathcal{UD}_{i+1, k}$ can be generated by grafting a leaf labeled with $i+1$ once (a), three times (b) and four times (c) from the dup-tree $M$ such that $L_2(M)=[i]$ and $M$ contains at most one twin-cherry.  Here, the circled subtrees consist of a duplication node and the associated  conjugate subtrees.
   ({\bf a}) The right-handed  unrooted dup-tree is in $\mathcal{UD}_{3, 5}$ that has two Leaves $3$ in the conjugate  subtrees of a duplication node. It can only be generated by grafting leaves with $3$ in a unique
edge in a unique dup-tree on the left.   ({\bf b}) None of the leaves labeled with 3 is in a conjugate subtree in the right-handed  dup-tree. Pruning the left-hand leaf labeled with 3 gives a dup-tree (left) that contains a new duplication node, but not for the right-handed  leaf labeled with 3.
Conversely, the right dup-tree can be generated from the left-handed dup-trees (in $\mathcal{UD}_{2, 5}$) by grafting a Leaf 3 in three ways.  ({\bf c}) Neither of the leaves labeled with  $4$ are in the conjugate subtrees of a duplication node in the right-handed  dup-tree. But pruning each of the leaves gives  a dup-tree (left) that contains a new duplication node.  Conversely, the right-handed  dup-tree can be generated from two left-handed dup-trees (in $\mathcal{UD}_{3, 5}$) by grafting a Leaf 4  four times.}
  \label{fig_4}
\end{figure}

By Lemma~\ref{lemma5}, a unrooted dup-tree in $\mathcal{UD}_{i+1, k}$ can be generated by grafting a leaf labeled with $i+1$ in a unique dup-tree twice, in two different dup-trees twice, in two dup-trees three or four times, as illustrated in 
Figure~\ref{fig_4}. 

We are now ready to establish  a formula for $|\mathcal{UD}_{i+1, k}|$. 
We will use the following parameters:
\begin{itemize}
   \item $\mathcal{C}_{i, k}$: the set of unrooted dup-trees $T$ over $[k]$ such that 
$L_2(T)=[i]$ and that contains only one twin-cherry. 
  \item $O_1$:  the number of unrooted dup-trees $T$ in $\mathcal{UD}_{i+1, k}$ such that 
  two leaves with $i+1$ are in the conjugate subtrees of a duplication node.

\item $O_3$: the number of unrooted dup-trees $T$ in $\mathcal{UD}_{i+1, k}$ such that
    the removal of one labeled with $i+1$ gives a dup-tree with one more duplication node than $T$, but 
  the removal of the other does not change the duplication nodes.

\item $O_4$: the number of unrooted dup-trees $T$ in $\mathcal{UD}_{i+1, k}$ such that
    the removal of either of  the leaves labeled with $i+1$ gives a dup-tree with one more duplication node than $T$.
\end{itemize}

To generate all the unrooted dup-trees in $\mathcal{UD}_{i+1, k}$, we graft 
another leaf labeled with $i+1$ in all but the edge incident to Leaf $i+1$ in each dup-tree in 
$\mathcal{UD}_{i, k}$
and in each edge of the unique twin-cherry in each dup-tree in $C_{i, k}$.
Since each dup-tree in $\mathcal{UD}_{i, k}$ has $k+i$ leaves and $2(k+i)-3$ edges,
we have the following identity: 
\begin{eqnarray}
 2(k+i-2) |\mathcal{UD}_{i, k}| + 2 |\mathcal{C}_{i, k}| =
  2|\mathcal{UD}_{i+1, k}| - O_1 + O_3 + 2O_4. \label{eqn6}
\end{eqnarray}

\begin{lemma} Let $i<k$. We then have:
\begin{eqnarray}
   && |\mathcal{C}_{i, k}|=i\cdot  |\mathcal{UD}_{i-1, k}|, \label{eqn7} \\
   && O_1=\sum_{1\leq d\leq i} {i \choose d}\cdot (2d-1)!! \cdot |\mathcal{UD}_{i-d, k-d}|, \label{eqn8} \\
  && O_3 +2O_4= \sum_{1\leq d\leq i} {i \choose d} \cdot  (2d-1)!! \cdot |\mathcal{UD}_{i-d, k-d+1}|. 
 \label{eqn9}
\end{eqnarray}
\end{lemma}
{\bf Proof.}
For any $S\subseteq [k]$ of $j$ elements,  the dup-trees $T$ over $[k]$ such that 
 $L_2(T)=S$ have one-to-one correspondence with the dup-trees $T'$ over $[k]$ such that 
 $L_{2}(T')=[j]$. 
The dup-trees in $\mathcal{C}_{i, k}$ that contains only one twin-cherry consisting of 
leaves labeled with $i'$ ($i'\leq i$) can be generated by grafting Leaf $i'$ in the unique edge incident to Leaf $i'$ in 
every twin-cherry-free dup-tree over $[k]$  such that $L_{2}=[i]-\{i'\}$.  Taken together, these two facts imply Eqn.~(\ref{eqn7}).

Note that $O_1$ is equal to the number of dup-trees over $[k]$ in which two Leaves $i+1$ appear in the conjugate subtrees of a duplication node.  We assume that $T$ is such a dup-tree over $[k]$ and $u$ is the duplication node whose conjugate subtrees contain leaves labeled with $j+1$. Let $T'=T\ominus \{\ell'(i+1), \ell''(i+1)\}$. $T'$ is then a dup-tree over $[k]-\{i+1\}$ such that 
$L_2(T')=[i]$ and $u$ remains as a duplication node in $T'$.

  Conversely, let $T''$ be a dup-tree over $[k]-\{i+1\}$ such that $L_2(T'')=[i]$. If
$T''$ contains a duplication node $u$ whose conjugate subtrees are of  $d$ leaves, then simultaneously grafting two 
leaves labeled with $i+1$ in each of the $(2d-2)$ pairs of conjugate edges in the conjugate subtrees of $u$ as well as in the two edges incident to $u$, we obtain $(2d-1)$ dup-trees over $[k]$ such that $u$ is still a duplication node.   Note that there are 
$(2d-3)!!$ rooted binary trees with $d$ leaves,  and  removing  the conjugate subtrees from $T''$ and  treating  $u$ as a leaf with a new label generates  a dup-tree $D$ with $(k-1)-d+1$ labels,  $i-d$ of which are duplicated labels.  Summing over all possible $d$ values  from $1$ to
$i$, we obtain:
 \begin{eqnarray*}
           O_1 & = & \sum_{1\leq d\leq i} \#\mbox{($d$-element subsets of $[i]$)}
            \cdot (1+\#\mbox{(edges in a tree in $\mathcal{T}_d$)}) \cdot 
           |\mathcal{T}_d| \cdot 
           |\mathcal{UD}_{i-d, k-d}|\\
   &=& \sum_{1\leq d\leq i} {i \choose d} (2d-1) \cdot (2d-3)!! \cdot  |\mathcal{UD}_{i-d, k-d}|.
\end{eqnarray*}
Therefore, Eqn.~(\ref{eqn8}) holds.

To prove Eqn.~(\ref{eqn9}), we let $\mathcal{P}_{i, k}(S)$ be the set of the unrooted 
dup-trees $T$ over $[k]$ in which $L_2(T)=[i]$ and where there is a duplication node $u$ whose conjugated subtrees have leaves  with labels   in $S$ for any $S\subseteq [i]$ of $d$ labels. Grafting a new leaf labeled with $i+1$, $\ell'(i+1)$,  in each edge in a fixed conjugate subtree of $u$ as well as an edge incident to $u$ gives 
$(2d-1)$ dup-trees $T'$ such that $T'\ominus \ell'(i+1)=T$. Clearly,  $T$ contains one more duplication node than 
such a $T'$.  Note that the removal of the conjugate subtrees of $u$ transforms each tree in $\mathcal{P}_{i, k}(S)$
into a tree $T''$ over $\{u\}\cup [k]-S$ such that $L_2(T'')=[i]-S$ if $u$ is considered to be a labeled leaf. Thus, 
$$|\mathcal{P}_{i, k}(S)|= (2d-1)\cdot (2d-3)!! \cdot  |\mathcal{UD}_{i-d, k-d+1}|.$$

Conversely, let $T \in \mathcal{UD}_{i+1, k}$. If $T\ominus \ell'(i+1)$ and $T\ominus \ell''(i+1)$ both contain a new duplication node compared with $T$, $T$ can be generated by grafting from two different dup-trees in 
$\cup _{S\subseteq [i]} \mathcal{P}_{i, k}(S)$. Therefore, 
\begin{eqnarray*}
  \sum_{S\subseteq [i]} |\mathcal{P}_{i, k}(S)|
 = \sum^{i}_{d=1} {i\choose d}\cdot (2d-1)!! \cdot |\mathcal{UD}_{i-d, k-d+1}|=O_3+2O_4.
\end{eqnarray*}
  This proves Eqn.~(\ref{eqn9}).
\QED

By plugging Eqn.~(\ref{eqn7})--(\ref{eqn9}) into Eqn.~(\ref{eqn6}), we obtain the following recursive formula for $|\mathcal{UD}_{i+1, k}|$.

\begin{theorem} 
Let $i < k$ and $N^{(i)}_{k}=|\mathcal{UD}_{i, k}|=|\mathcal{G}_{i, k-1}|$. We then have: 
 \begin{eqnarray}
N^{(i+1)}_{k} = (k+i-2) N^{(i)}_{k} + i  N^{(i-1)}_{k}
  +  \frac{1}{2}\sum_{1\leq d\leq i} {i \choose d} (2d-1)!! \left( N^{(i-d)}_{k-d}  - N^{(i-d)}_{k-d+1} \right).
    \label{eqn10}
\end{eqnarray}
\end{theorem}

\noindent {\bf Example \ref{sect5}.1} For $k=4$, we have:
\begin{eqnarray*}
  &&N^{(0)}_{4}=(2\times 4 - 5)!! = 3,\\
  &&N^{(1)}_{4}=(4-2)N^{(0)}_{4}=6,\\
  && N^{(2)}_{4}=3 N^{(1)}_{4} +  N^{(0)}_{4} + \frac{1}{2}\left(N^{(0)}_{3} - N^{(0)}_{4}\right)
  = 20,\\
  && N^{(3)}_4=4N^{(2)}_{4}+2 N^{(1)}_{4} + \frac{1}{2}\left[ {2 \choose 1} 
        \left(N^{(1)}_{3} - N^{(1)}_{4}\right) + {2\choose 2} 3!!  \left(N^{(0)}_{2} - N^{(0)}_{3}\right)\right]=87.\\
\end{eqnarray*}

Table~\ref{table1} lists the values of $N^{(i)}_{k}$ for  $i$ and $k$ such that  $0\leq i <k$ and $2\leq k\leq 10$.

\begin{table}[t!]
\caption{The values of $N^{(i)}_{k}$ for $0\leq i<k$ and $2\leq k\leq 10$. \label{table1}}
  \centering 
   \includegraphics[scale = 0.6]{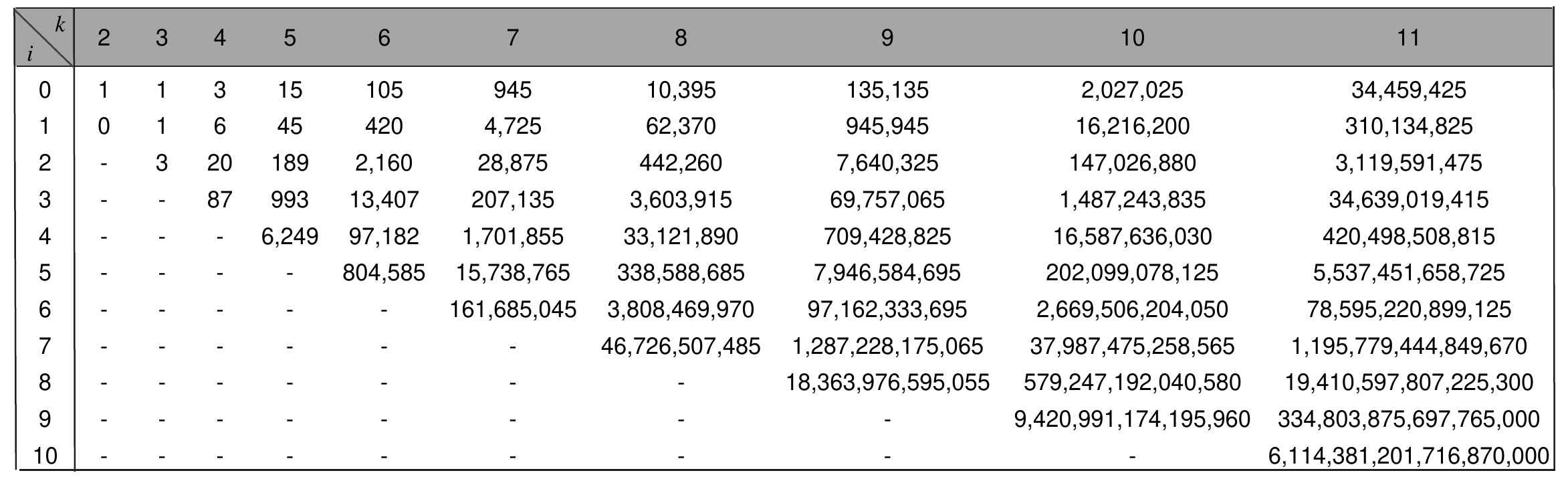}
\end{table}

\subsection{A formula for counting 1-galled networks}

A 1-galled network over $[k]$ may contain 0 to $k$ reticulate nodes.
Since  the 1-galled networks with $i$ reticulate nodes have one-to-one correspondence with the rooted 
dup-trees over $[k]$  that have $i$ duplicated labels, they have one-to-one correspondence with 
the unrooted dup-trees over $[k+1]$ that have $i$ duplicated labels. Therefore,  we have the following theorem.

\begin{theorem}
\label{thm5.2}
Let $G_1(k)$ denote the number of 1-galled networks over $k$ taxa. We then have,    
\begin{eqnarray}
G_1(k)=\sum^{k}_{i=0} {k \choose i} N^{(i)}_{k+1}, \label{eqn11}
\end{eqnarray}
where 
$N^{(i)}_{k+1}$ is defined in Eqn.~(\ref{eqn10}).
\end{theorem}

\noindent {\bf Example \ref{sect5}.1} (con't) By Theorem~\ref{thm5.2}, the number of 1-galled network on three taxa is:  $$N^{(0)}_4 + {3\choose 1} N^{(1)}_{4} + {3\choose 2} N^{(2)}_{4}+ {3\choose 3} N^{(3)}_{4}
=168.$$ 
All 34 topological structures of these 168 1-galled networks are drawn in Figure~S1.

\section{Counting general galled networks}
\label{sec5}

\begin{figure}[b!]
            \centering
            \includegraphics[scale = 0.8]{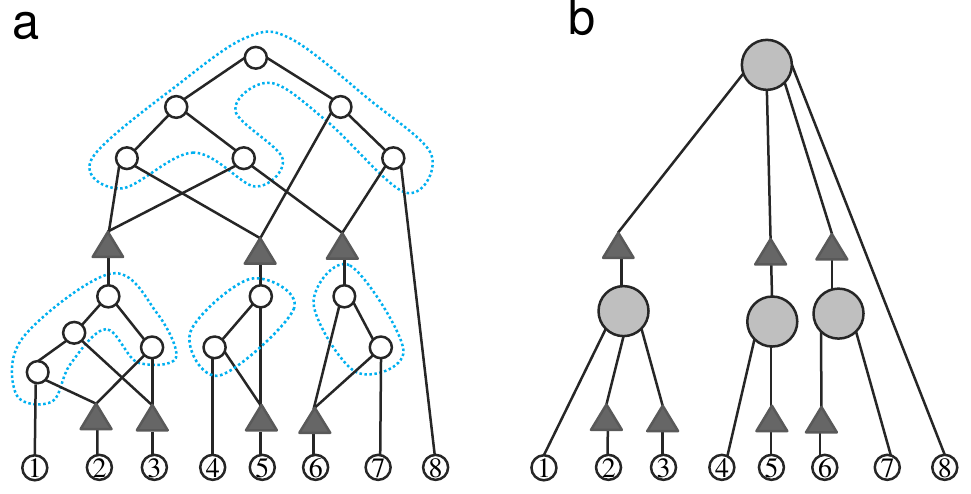}
  \caption{Illustration of the network compression operation. ({\bf a})  A galled network over $[8]$. It has four tree-components.
({\bf b}) The compression of the network in (a). It is a rooted tree in which each reticulate node becomes a node of indegree 1 and outdegree 1.}
\label{Fig6}
\end{figure}

\subsection{Compression of galled networks}

The technique of network decomposition was first introduced to study two algorithmic problems for RPNs in \cite{Gunawan_16}.  Recently,   component-wise compression was formally investigated to reveal the connection between several classes of RPNs \cite{Gunawan_18}. Intuitively, \textit{compressing} a RPN $N$ involves  replacing every component in $N$ with a node of degree 2 or more, thereby creating a smaller network $\Tilde{N}$ that summarizes the relationships among tree-components in $N$. The node and edge sets of the compression network $\Tilde{N}$ of $N$ is rigorously defined as follows:
    \begin{align*}
        \mathcal{V}(\Tilde{N}) = & \mathcal{L}(N) \cup \{v_\tau: \tau \text{ is a tree- or reticulation component in } N\}, \text{ and}\\
        \mathcal{E}(\Tilde{N}) = & \{(v_\tau,\ell): \ell \in \mathcal{L}(N)\text{ and } p(\ell) \in \tau\} \\ &\cup  \{(v_\tau,v_\tau'): \text{there is } (x,y)\in \mathcal{E}(N) \text{ such that } x\in \tau, y \in \tau'\}, 
    \end{align*}
    where $p(\ell)$ denotes the parent of the leaf $\ell$. 
The operation of network compression is illustrated in Figure~\ref{Fig6}.

 In a galled network,  each reticulate node is inner and thus both its parents are in a common tree-component.  Therefore,  a tree-component becomes  a node with at least two children and each reticulate node becomes  a node of indegree 1 and outdegree 1 after the tree-components are compressed. Thus,  the compression of a galled network is a tree (Theorem 3.1, \cite{Gunawan_18}) (see Figure~\ref{Fig6}), 
implying that a galled network consists of a set of 1-galled networks stacked one on the top of the other in a tree shape.

\subsection{A counting method}

 We are now ready to count general galled networks over $[k]$. 
Let $\mathcal{A}_k$ be the set of non-binary phylogenetic trees over $[k]$ in which  every non-leaf node has two or more children.   Assume that $T \in \mathcal{A}_k$. For a non-leaf node $v \in \mathcal{V}(T)$, we use $c_{\rm lf}(v)$ and $c_{\rm nlf}(v)$ to denote  the numbers of  leaf  and non-leaf children of $v$ in $T$, respectively, and define
 $c(v)=c_{\rm lf}(v) + c_{\rm nlf}(v)$. Clearly $c(v)$ is the number of the children of $v$ in $T$.

Consider  a binary galled network $N$ over $[k]$.  By Theorem 3.1 in \cite{Gunawan_18}, 
the compression $C(N)$ of $N$ is a tree over $[k]$. A node of indegree and outdegree 1 in $C(N)$ corresponds one-to-one to a reticulate node in $N$, whereas  a tree node with two or more children in $C(N)$ corresponds one-to-one to a tree-component in $N$. 
For convenience,  we suppress all the nodes of indegree and outdegree 1 in $C(N)$ to get  rooted  tree $C'(N) \in \mathcal{A}_k$. Clearly, the tree-components of $N$ are still in one-to-one correspondence with the tree nodes in $C'(N)$. By reverse-engineering this process, we can enumerate and count general galled networks over $[k]$, as all possible general rooted trees over $[k]$ can be enumerated and counted recursively \cite{Felsenstein}. 
    
    \begin{theorem}
\label{thm6.1}
    Let $G(n)$ be the number of galled networks over $n$ taxa. We then have:
        \begin{equation}
            G(n) = \sum_{T \in \mathcal{A}_k}
            \left(\prod_{v \in \mathcal{I}(T)} \left(\sum^{c(v)}_{j=c_{\rm nlf}(v)}  \binom{c_{\rm lf}(v)}{c(v)-j} 
  N^{(j)}_{c(v)+1}\right) \right),
        \end{equation}
where $\mathcal{I}(T)$ denotes the set of non-leaf nodes in $T$ and $N^{(j)}_{c(v)+1}$ is defined in Eqn.~(\ref{eqn10}).
    \end{theorem}
{\bf Proof.}  Let $T\in \mathcal{A}_k$ such that $C'(N)=T$ for some galled network $N$. Consider a non-leaf node $v$ in $T$. We first consider how to reconstruct the tree-component $\sigma$ of $N$ that corresponds to $v$. 
 Let $R$ denote the set of reticulate nodes in $N$ whose parents are both in $\sigma$. For each child $u$ of $v$ that is a non-leaf, the root of the tree-component corresponding to $u$ must be a child of a reticulate node in $R$. However, for each child $\ell$ of $v$ that is a leaf, the parent of $\ell$ in $N$ may be a tree node in $\sigma$ or a reticulate node in $R$. Therefore, $ c_{\rm nlf} (v) \leq |R| \leq c(v)$ and $R$ has ${c_{\rm lf} \choose |R|-c_{\rm nlf}(v)}$ possibilities. For each possible selection 
of $R$, $\sigma$ corresponds to a 1-galled network  with $c(v)$ leaves and $|R|$ reticulate nodes. Thus,  the component  $\sigma$ corresponding to $v$ has $\sum^{c(v)}_{j=c_{\rm nlf}(v)}  \binom{c_{\rm lf}(v)}{c(v)-j} 
  N^{(j)}_{c(v)+1}$ choices. 
  
Since the reconstructions of two distinct tree-components in $N$ are independent from each other, the number of galled networks whose compression correspond to $T$ is \\
$\prod_{v \in \mathcal{I}(T)} \left(\sum^{c(v)}_{j=c_{\rm nlf}(v)}  \binom{c_{\rm lf}(v)}{c(v)-j} 
  N^{(j)}_{c(v)+1}\right)$. Hence, the theorem follows.
  \QED

The numbers $G(n)$ of galled networks over $n$ taxa was calculated according to Therem~\ref{thm6.1} and are listed in 
Table~\ref{table2}. For example, $G(3)=240$. This implies that  there are $240-168=72$ galled networks with two tree-components over three taxa,  the topological structures of which are listed in Figure~\ref{figs2}. 
    
    \begin{table}[h!]
 \caption{The values of $G(n)$ for $1\leq n\leq 10$. \label{table2}}
        \centering
        \includegraphics{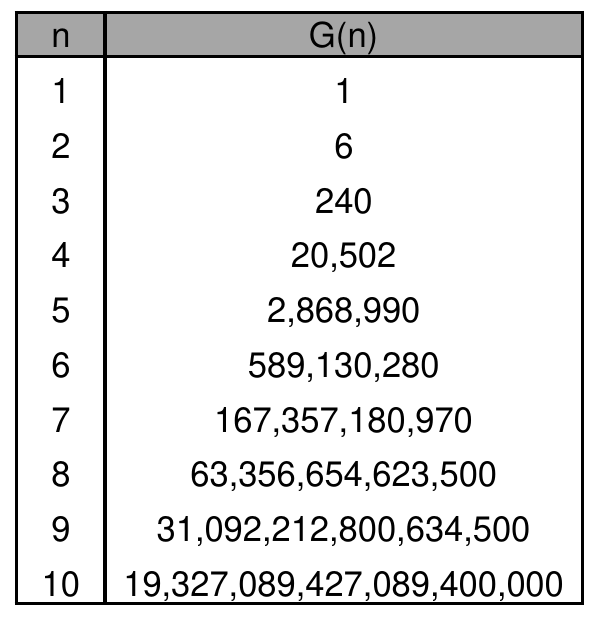}
    \end{table}

\section{Conclusion}
\label{sec6}

We have presented a linear recurrence formula for counting all possible 1-galled networks and a method for counting and enumerating general galled networks. We conclude the study with a couple of remarks.

First, using the same counting technique as in Section 4.1, we can derive the following recurrence formula for the number 
of all  unrooted dup-trees $T$ such that $L_2[T]=[i]$, denoted by $B^{(i)}_{k}$:
  \begin{eqnarray}
     &&B^{(0)}_{k}=(2k-5)!!, \nonumber \\
     && B^{(1)}=(k-1)\cdot (2k-5)!! \nonumber\\
     && B^{(i+1)}_{k}= (n+k-1) B^{(i)}_{k} + \frac{1}{2}\sum_{1\leq d\leq i}{i \choose d}\cdot (2d-1)!!\cdot \left(B^{(i-d)}_{k-d} - B^{(i-d)}_{k-d+1} \right).
\end{eqnarray}

Second,  
 galled networks form a subclass of  reticulation-visible networks. We therefore pose counting  reticulaiton-visible networks as an open question.

\section*{Acknowledgements}

The authors thank HW Yan and Jonathan M. Woenardi for participanting in discussion on this work. 
This work was supported by Singapore Ministry of Education Academic
Research Fund Tier-1 [grant R-146-000-238-114] and National Research
Fund [grant NRF2016NRF-NSFC001-026].

\newpage
\setcounter{figure}{0}
\renewcommand{\thefigure}{S\arabic{figure}}

\begin{figure}[h!]
            \centering
            \includegraphics[scale = 0.8]{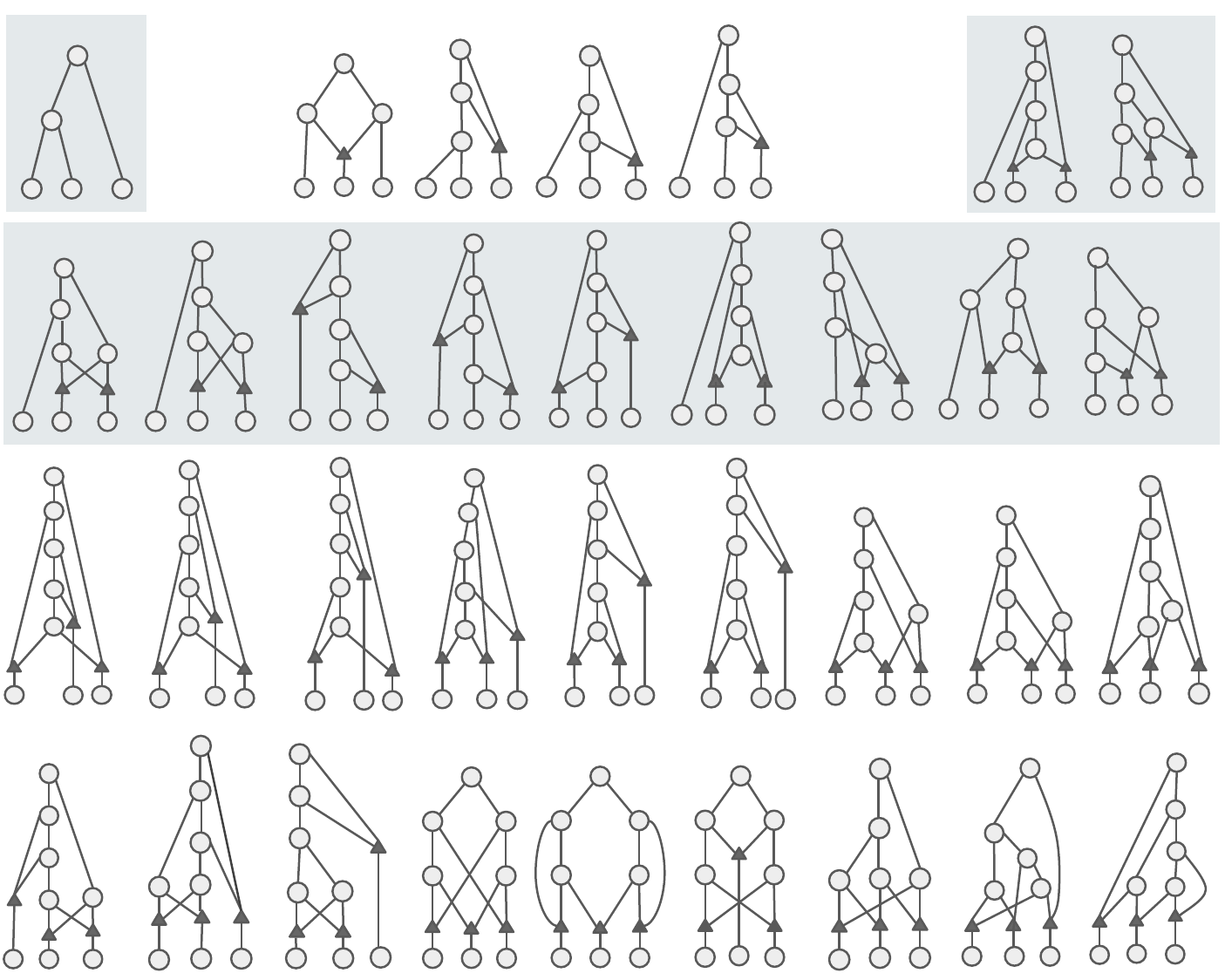}
  \caption{The 34 toplogical structures of the 168 1-galled networks over three taxa.}
\label{fig5}
\end{figure}

\begin{figure}[h!]
            \centering
            \includegraphics[scale = 0.8]{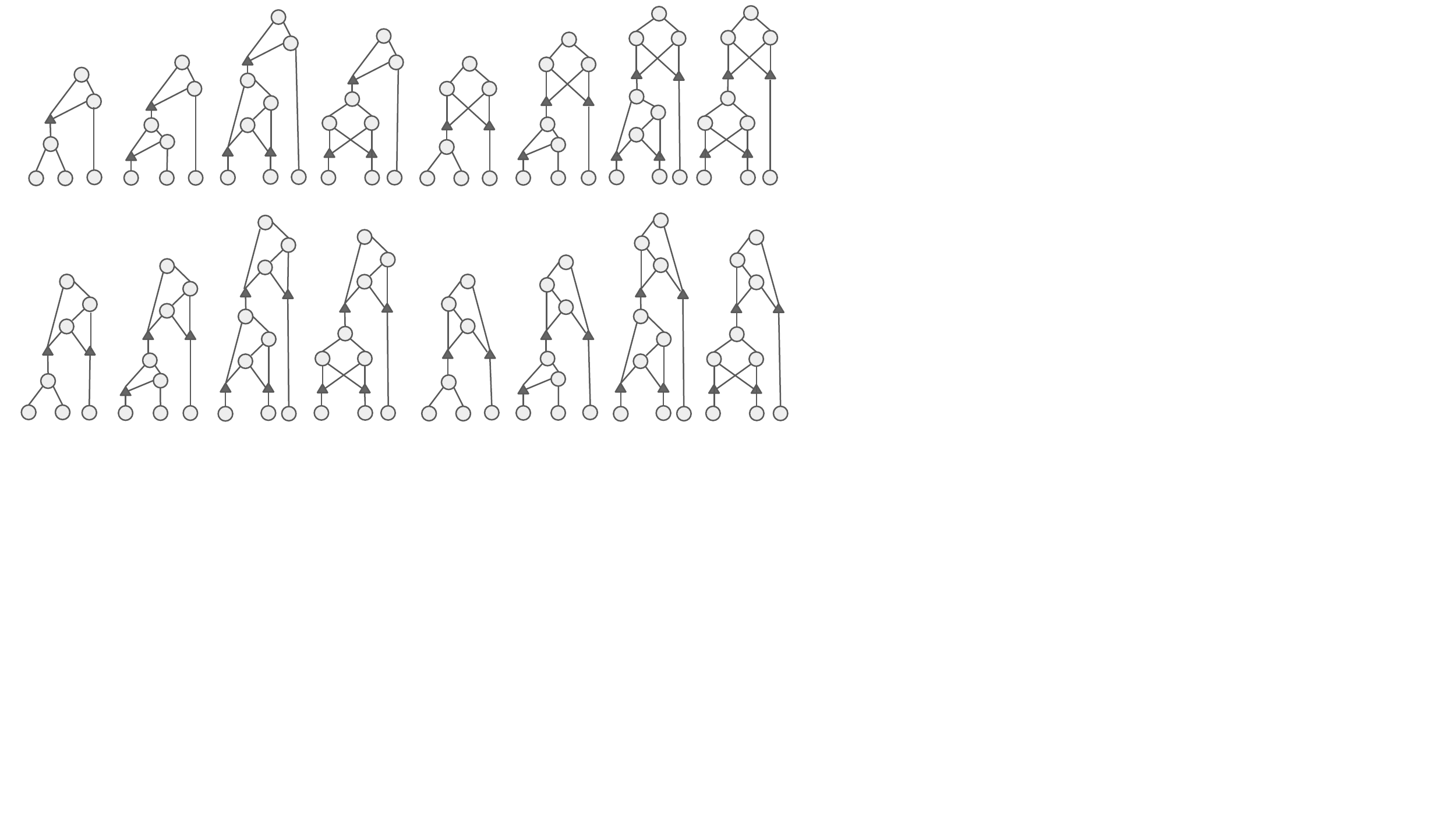}
  \caption{The 16 toplogical structures of the 72 galled networks over three taxa that have two tree-components.}
\label{figs2}
\end{figure}


\begin{thebibliography}{10}
\providecommand{\url}[1]{{#1}}
\providecommand{\urlprefix}{URL }
\expandafter\ifx\csname urlstyle\endcsname\relax
  \providecommand{\doi}[1]{DOI~\discretionary{}{}{}#1}\else
  \providecommand{\doi}{DOI~\discretionary{}{}{}\begingroup
  \urlstyle{rm}\Url}\fi





\bibitem{Bordewich_16}
Bordewich, M., Semple, C.: Reticulation-visible networks.
\newblock Adv. Applied Math. \textbf{78}, 114--141 (2016)

\bibitem{Bouvel_18}
Bouvel. M., Gambette, P., Mansouri, M.: Counting level-k phylogenetic networks. 
In preparation (2018)

\bibitem{Cardona_09}
Cardona, G., Llabr{\'e}s, M., Rossell{\'o}, F., Valiente, G.: Metrics for
  phylogenetic networks i: Generalizations of the Robinson-Foulds metric.
\newblock IEEE/ACM Trans. Comput. Biol. Bioinform. \textbf{6}(1), 46--61 (2009)

\bibitem{Cardona_09b}
Cardona, G., Rossello, F., Valiente, G.: Comparison of tree-child phylogenetic
  networks.
\newblock IEEE/ACM Trans. Comput. Biol. Bioinform. \textbf{6}(4), 552--569
  (2009)

\bibitem{Moulton_13}
Czabarka {\'E}, Erd{\H{o}}s PL, Johnson V., Moulton V.: Generating functions for multi-labeled trees. 
Discrete  Applied Math. 161, 107-117 (2013)

\bibitem{Felsenstein}
 Felenstein, J.:  Inferring Phylogenies. Sinauer Associates,   Sunderland, MA, USA (2004)


\bibitem{Francis_15}
Francis, A.R., Steel, M.: Which phylogenetic networks are merely trees with
  additional arcs?
\newblock Syst. Biol. \textbf{64}(5), 768--777 (2015)

\bibitem{Fuchs_18}
Fuchs, M., Gittenberger, B. and Mansouri, M.:  Counting phylogenetic networks with few reticulation vertices: Tree-child and normal networks. arXiv preprint arXiv:1803.11325 (2018)

\bibitem{Gambette_15}
Gambette, P., Gunawan, A.D., Labarre, A., Vialette, S., Zhang, L.: Locating a
  tree in a phylogenetic network in quadratic time.
\newblock In: Proc. Int'l Confer. on Res. in Comput. Mol. Biol. (RECOMB), pp.
  96--107. Springer, New York (2015)





\bibitem{Gunawan_16}
Gunawan, A.D., DasGupta, B., Zhang, L.: A decomposition theorem and two
  algorithms for reticulation-visible networks.
\newblock Inform. Comput. \textbf{252}, 161--175 (2017)

\bibitem{Gunawan_18}
Gunawan, A.D., Yan, H., Zhang, L.: Compression of phylogenetic networks and algorithm
for the tree containment problem.
J. Comput. Biol. (in press). 
ArXiv preprint arXiv:1806.07625 (2018)

\bibitem{Gusfield_book}
Gusfield, D.: ReCombinatorics: the Algorithmics of Ancestral Recombination
  Graphs and Explicit Phylogenetic Networks.
\newblock MIT Press, Boston, USA (2014)

\bibitem{Gusfield_04}
Gusfield, D., Eddhu, S., Langley, C.: The fine structure of galls in
  phylogenetic networks.
\newblock INFORMS J. Comput. \textbf{16}(4), 459--469 (2004)

\bibitem{Huber_06}
Huber KT, Moulton V.:
 Phylogenetic networks from multi-labelled trees. J. Math. Biol. 52, 613--632 (2006)

\bibitem{Huson_07}
Huson, D.H., Kl{\"o}pper, T.H.: Beyond galled trees--decomposition and
  computation of galled networks.
\newblock In: Proc. Int'l Confer. on Res. in Comput. Mol. Biol. (RECOMB), pp.
  211--225. Springer, New York, USA (2007)

\bibitem{Huson_09}
Huson, D.H., Rupp, R., Berry, V., Gambette, P., Paul, C.: Computing galled
  networks from real data.
\newblock Bioinformatics \textbf{25}(12), i85--i93 (2009)

\bibitem{Huson_book}
Huson, D.H., Rupp, R., Scornavacca, C.: Phylogenetic networks: Concepts,
  Algorithms and Applications.
\newblock Cambridge University Press, Cambridge, UK (2010)

\bibitem{Lake_99}
Jain R, Rivera MC, Lake JA.:  Horizontal gene transfer among genomes: the complexity hypothesis. 
Proc. Nat'l Acad. Sc. U.S.A. 96, 3801--3806 (1999)

\bibitem{Semple_15}
McDiarmid, C., Semple, C. and Welsh, D.: Counting phylogenetic networks. Annals Combin. 19, 205--224 (2015)

\bibitem{Steel_06}
Semple, C. and Steel, M.:  Unicyclic networks: compatibility and enumeration. IEEE/ACM Trans. Comput. Biol. Bioinform. 3, 84--91 (2006)

\bibitem{Steel_book}
Steel, M.: Phylogeny: Discrete and Random Processes in Evolution.
\newblock SIAM, Philadelphia, USA (2016)





\bibitem{Wang_01}
Wang, L., Zhang, K., Zhang, L.: Perfect phylogenetic networks with
  recombination.
\newblock J. Comput. Biol. \textbf{8}(1), 69--78 (2001)


\bibitem{Gunawan_17}
Yan, H., Gunawan, A.D., Zhang, L.: S-cluster++: a fast program for solving the
 cluster containment problem for phylogenetic networks.
\newblock Bioinformatics \textbf{34}(17), i680--i686

\bibitem{Zhang_16}
Zhang, L.: On tree-based phylogenetic networks.
\newblock J. Comput. Biol. \textbf{23}(7), 553--565 (2016)

\bibitem{Zhang_18}
Zhang, L.: Clusters, trees and phylogenetic network
classes. 
\newblock In T. Warnow (ed.):  Bioinformatics and Phylogenetics: Seminal Contributions of Bernard Moret,  Springer, New York 
(2019)


\end{thebibliography}
\end{document}